\documentclass[multphys,vecphys]{svmult}

\def\solm{M$_{\odot}\,$}

\def\kms{km s$^{-1}$}

\usepackage{makeidx}         
\usepackage{graphicx}        
\usepackage{multicol}        
\usepackage[bottom]{footmisc}

\makeindex             

\begin{document}

\title*{The History of Galaxy Formation in Groups: An Observational Perspective}
\titlerunning{Galaxy Evolution in Groups}
\author{Christopher J. Conselice}
\institute{School of Physics and Astronomy, University of Nottingham, UK}
%
%
\maketitle

{\bf Summary.} We present a pedagogical review on the formation and
evolution of galaxies in groups, utilizing observational 
information from the Local Group to
galaxies at $z \sim 6$.  The majority of galaxies in the nearby
universe are found in groups, and galaxies at all redshifts up to
$z \sim 6$ tend to cluster on the scale of nearby groups ($\sim$1 Mpc). This 
suggests that the group environment may play a role in the formation of 
most galaxies. The Local Group, and other nearby groups, display a diversity 
in star formation and morphological properties that puts limits on how, 
and when,
galaxies in groups formed.  Effects that depend on an intragroup medium,
such as ram-pressure and strangulation, are likely not major mechanisms
driving group galaxy evolution.   Simple dynamical friction arguments 
however show that galaxy mergers should be common, and a dominant process 
for driving evolution. While mergers between L$_{*}$ galaxies are 
observed to be rare at $z < 1$, they are much more common at earlier times.  
This is due to the increased density of the universe, and to the fact that 
high mass galaxies are highly clustered on the scale of groups.  
We furthermore discus why the {\em local} number
density environment of galaxies strongly correlates with galaxy properties,
and why the group environment may be the preferred method for establishing 
the relationship between properties of galaxies and their local density.   

\section{Introduction}
\label{sec:1}

Astronomers have known  since the time of
Messier and the Herschels that the faint nebula, or what we today call
galaxies, cluster together. This was before we knew
anything else about these systems, including their distances. Galaxy 
clustering remains one of the cornerstones of cosmology and galaxy formation,
and most galaxies are clustered in some form.  This is one of the major
successes of the Cold Dark Matter model of structure formation, and
simulations show that the bulk of large-scale
structure is composed of individual groups of galaxies [37]. This
is found to be the case observationally [24], and it appears that up to half of all nearby galaxies are 
in groups or clusters [26].

The fact that a significant fraction of all galaxies are found in groups is 
likely an important aspect for understanding galaxy 
formation and evolution.  We know that the local environment\footnote{We often
refer to local and global environments in this paper.  The 
local environment is the density defined by the volume enclosing
a galaxy and its nearest bright neighbors. The global environment refers to
the type of environment a galaxy lives, whether it be a cluster, a group, or 
the field, without regards to whether the galaxy is in the core or outer
part of a cluster or group.  The local environment can be quantitatively 
measured through
a nearest {\em N} neighbor approach detailed in e.g., [23,50], or through a 
friends-of-friend algorithm. }  of a 
galaxy correlates with most of its properties.  
The most famous example of this is the 
morphology-density relation [23], where galaxies with
early-type morphologies are more likely found in denser areas, while 
spirals are more likely found in lower density environments.   Galaxies in 
low local density regions also have a higher star formation rate than
those in areas with a higher local galaxy density [40,30]. 
Whether the relationship between local density and the properties of galaxies
is intrinsic, or is a result of physical processes that occur in dense regions 
after initial galaxy formation is still an unresolved issue.

Galaxy evolutionary effects were also first noticed in the densest areas of 
the universe - namely galaxy clusters.  A high
fraction of blue galaxies were found in galaxy clusters 
at $z \sim 0.5$ compared to local systems [4].
These star forming, or post-star forming galaxies, possibly provide
evidence that denser environments induce relatively recent evolution in 
galaxies.
In fact, it is largely unarguable that dense environments such as 
groups\footnote{We define a group as a gravitationally bound
system of galaxies with less than fifty members all within 1-2 Mpc, and
with only a few bright $> L_{*}$ members. Typically,
these systems have masses $\sim 10^{13} - 10^{14}$~\solm, and
velocity dispersions of 150-500 \kms.} and
clusters induce {\em some} evolution. When these effects occur, and
to what degree they alter the evolution of galaxies, is still open for debate.

In this review, we address some of these issues by focusing on the 
most common environment of galaxies - the
galaxy group - and how the cumulative effects of a dense galaxy
environment drives the evolution and formation of its members.
There are several reasons why the group environment is perhaps the most 
important for understanding how galaxy formation and evolution occurs.
The main reason is due to the fact that most nearby galaxies are 
in group environments, which
we now know extends up to $z \sim 1.4$ [27].  Furthermore, 
many of the physical processes associated with galaxy formation, e.g.,
galaxy mergers, can {\em only} occur in group-like environments.
Therefore groups appear to be a gateway environment producing 
galaxies with drastically different morphological and star forming
properties from previous field galaxies.

The outline of this review is as follows: we first describe the 
final (thus far) evolution of groups of galaxies and their properties by
examining nearby groups. We then extended these results to higher
redshifts and examine processes, such as star formation and galaxy mergers,
and how these might be driven by the group environment. Finally, we draw
some conclusions regarding how the group environment might be influencing the
evolution of most galaxies.  The cosmology H$_{0}$~=~70~km~s$^{-1}$~Mpc$^{-1}$,
${\rm \Omega_{\lambda}}$ = 0.7, and ${\rm \Omega_{m}}$ = 0.3 is used throughout.

\section{Groups in the Nearby Universe}
\label{sec:2}

\subsection{The Local Group}

The nearby universe provides a number of important clues for understanding how
galaxies in groups evolved.  The best studied example is of course our
own Local Group, with its 35+ members, each of which has undergone a 
distinct formation history (see [31] and the Grebel review
in these proceedings).  The most effective way to 
determine when stars in Local Group galaxies formed is to study the
resolved stellar populations within these systems.  The derived star formation
history of the Local Group is clearly very extended and variable, even
for the most basic dwarf spheroidals.  The star formation in these
simple and morphologically indistinct galaxies occurred from roughly
the time of reionization until a few Gyr ago [32].   
Dwarf irregulars, and the spirals in the Local Group, including our own 
Milky Way, are by definition still undergoing star formation, but
they also have old stars produced likely before reionization.

There are a few more interesting facts about the Local Group worth
mentioning in regards to the history of galaxy evolution.  While the Local
Group contains four massive members (M31, the Milky Way, M33 and the LMC), it
is dominated by lower mass dwarf galaxies.  These dwarf galaxies tend to
cluster around the two massive members, M31 and the Milky Way.  There is
also a strong environmental effect occurring within the Local Group dwarf
population.   The evolved
dwarf spheroidal galaxies are located nearest the two giant spirals, 
while the star forming dwarf irregulars are located away from the
giants. This is one example of how the {\em local} environment of a galaxy 
correlates with its properties.  For more extensive reviews of
the star formation history and properties of Local Group galaxies see
[31,53].

\subsection{Nearby Groups and the Morphology-Density Relation}

Nearby groups, such as the Sculptor and M66 groups (Figure~1), contain a 
similar range of morphological and ongoing star formation properties as the 
Local
Group.  Because of their distances, it is difficult to reconstruct
detailed star formation histories in these nearby groups,
but they likely have similar histories as the Local Group.

\begin{figure}
\centering
\includegraphics[height=6cm]{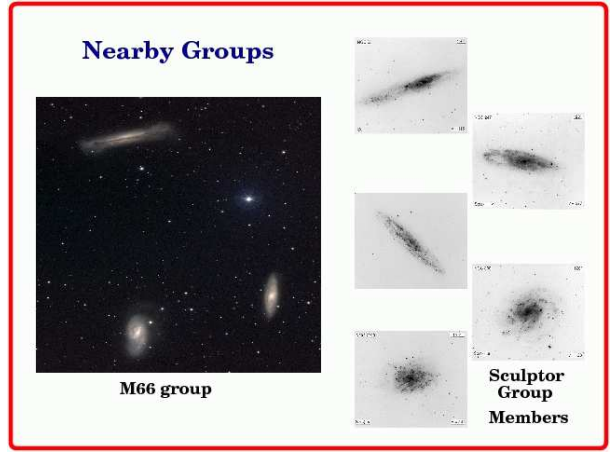}
%
%
\caption{Examples of galaxies in two of the nearest galaxy groups - the
M66 group and the Sculptor group. The galaxies in these groups are
similar to the Local Group, with many galaxies undergoing star formation,
and in the M66 groups an ongoing interaction between members.}
\label{fig:1}       
\end{figure}

We can still however use other tools, such as morphologies
and ongoing star formation in nearby group members, to understand how
these systems are evolving.  Star
formation is still occurring in both large and small galaxies in group environments.
Unless these groups recently formed, this implies that environmental effects 
that can reduce star formation, such as
ram-pressure stripping [44] or the gradual 
depletion of hot halo gas around galaxies (so-called strangulation or
starvation) [38], are not occurring - or at the very least they are not 
dominating effects. However, we can see environmental effects inducing 
evolution in the
form of induced star formation, and morphological distortions
due to galaxy-galaxy interactions, such as between M65 and M66, in the 
M66 group (Figure~1).

The morphology-density relation for nearby and distant groups and
clusters reveals that galaxy formation is sensitive to local
environment.  This correlation is such that the higher the local projected
galaxy surface density, the less likely a galaxy in that area will be a 
spiral, or 
undergoing active star formation [49]. However, 
[55] showed
that the morphological-density relation does not hold in a global
sense.  Zabludoff \& Mulchaey [55] found that groups of galaxies can have 
total early-type fractions as low as in the field, or as high as in clusters 
(fraction $\sim 0.6$), which is independent of the global environment.
The local environmental density in which group early-type galaxies are 
found is as high as the densest environments in clusters of 
galaxies. The passive or early type galaxies in groups are also found
in the centers of groups, where the local galaxy density
is highest.   

The star formation rate of a galaxy also does not correlate 
strongly, if at all, with its global environmental density
as measured by velocity dispersions [3].  
It is not the global environment, such as living in a massive cluster, but
the local environment which correlates with galaxy properties.  Likewise,
gravitational interactions between nearby galaxies only alter morphology,
and induce star formation, when they are separated by less than a galaxy 
diameter [34].  Galaxy
formation is therefore a local process. 

Another nearby galaxy group type that deserve detailed discussion are
the so-called compact groups. The compact groups, such as Seyfert's Sextent
and Stephen's Quintet, are examples of galaxy groups where the members
are within a galaxy diameter, and are likely to merge within about a Gyr.
While there is some controversy over the existence of compact groups - some
have argued they are chance alignments - the fact that there is a hot
intragroup medium associated with these galaxies provides strong evidence 
that they are bound objects.

An indication that compact groups may be the progenitors of mergers 
between galaxies are the fossil groups [48], ghost groups,
and the AWM groups [1].  
These objects are all systems with massive and luminous X-ray 
emission, but only contain one bright central galaxy.  There are some 
differences, such as the AWM group's central galaxy having a cD like 
structure.  However, what is clear about these `groups' is that although
they consist of only one bright galaxy, they have X-ray profiles and dark 
matter halos that closely resemble groups. This suggests that these systems
are recent merger remnants or the final stages of a galaxy group whose
members merged together.

\subsection{Galaxy Groups up to $z \sim 1.4$}
\label{sec:3}

Beyond about $z \sim 0.3$ there have been few
searches and systematic studies of groups of galaxies, although
this is rapidly changing. The first
evidence that galaxy groups at high redshift evolve was provided by 
Allington-Smith et al. [1] who discovered populations of galaxies
surrounding radio sources at $z \sim 0.5$. They furthermore found that
at the same local environmental density, there is a larger amount of star 
formation in galaxies in groups at $z \sim 0.3$ compared to lower redshift 
clusters.  This is similar 
to the Butcher-Oemler effect found in clusters, and possibly arises from
the same mechanism(s).

The first proper redshift surveys at $z > 0.3$ found significant peaks in 
galaxy redshift distributions [10,7], suggesting that real over-densities of galaxies exist beyond 
the local universe.  The CNOC2 survey pioneered efforts to characterize the 
galaxy population in groups at these redshifts [54], 
which has
only recently been superseded by the DEEP2 and VVDS redshift surveys 
[21,27,36].

The CNOC2 groups at redshift $z \sim 0.5$ display remarkably similar
scaling properties as field galaxies at similar redshifts, and in
comparison to $z \sim 0$ 
groups. [54] found a large blue galaxy fraction in both the 
field and in groups up to $z \sim 0.5$, with a similar rate of decrease in
both environments at lower redshifts.  
However, there are more passive galaxies in group environments
at all redshifts, suggesting that the correlation of environment in
groups with the properties of its member galaxies existed at least 5-6 Gyr ago.

\begin{figure}
\centering
\hspace{-0cm}
\includegraphics[height=12cm, angle=90]{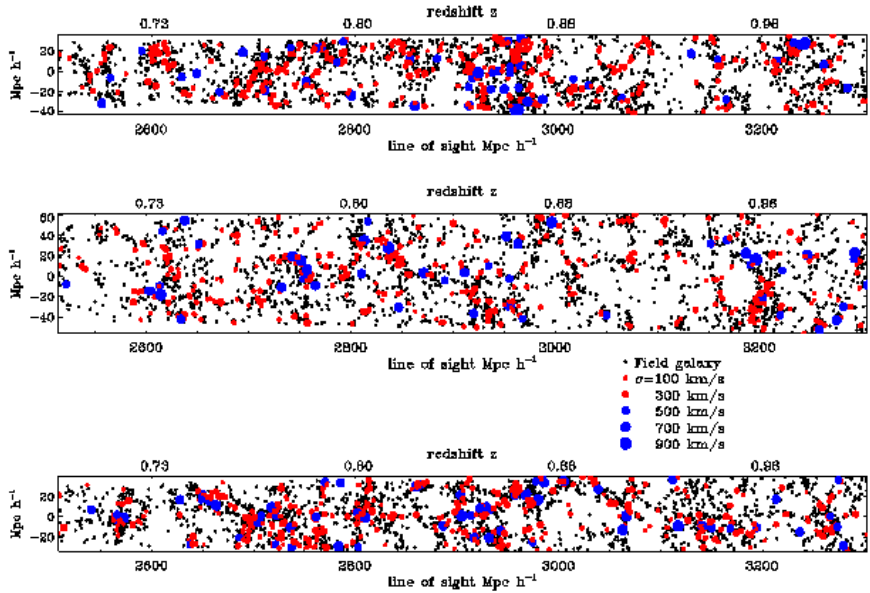}
%
%
\caption{The distribution of groups in redshift space as seen in the
DEEP2 redshift survey [27].  The larger symbols represent
groups/clusters with larger velocity dispersions, with groups
at dispersions of $z < 300$ \kms\, labelled as the smaller (and red) 
points. (Courtesy of Brian Gerke)}
\label{fig:1}       
\end{figure}

Redshift surveys with a high velocity resolution, such as DEEP2 and the VVDS,
are allowing
us to trace how galaxies cluster out to $z \sim 1$.  One result of
these studies is that galaxies, particularly bright galaxies, cluster 
strongly out to $z \sim 1.4$ [9].  Using various 
cluster/group finding techniques such as the Voronoi-Delaunay method,
individual groups of galaxies can be identified and studied
out to these redshifts [27] (Figure~2).

\begin{figure}
\centering
\hspace{-1.15cm}
\includegraphics[height=8cm]{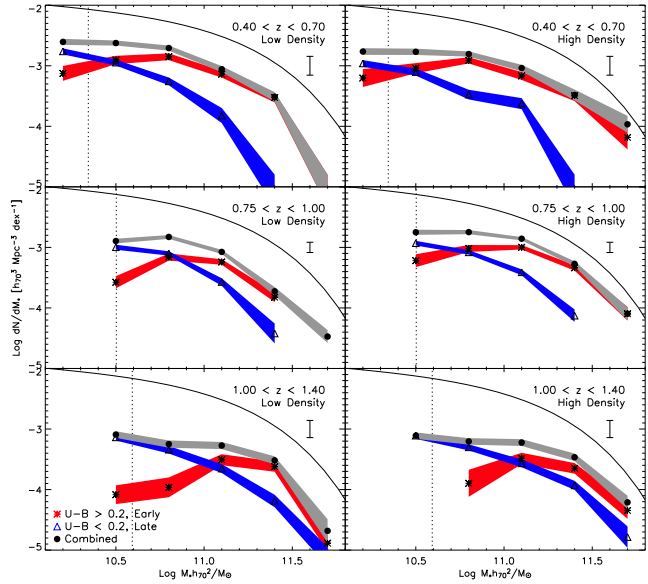}
%
%
\caption{The stellar mass function from [5] plotted
as a function of density and redshift (time).  The mass functions
are furthermore divided into red and blue galaxies.  The transition between 
blue and red galaxy dominance, i.e., when M$_{\rm blue} >$ M$_{\rm red}$ 
does not change with a broad environmental density cut.}
\label{fig:1}       
\end{figure}

Once we have found these groups we can then try to determine how 
environment drives evolution up to $z \sim 1.4$.  Bundy et al. [5] have
approached this
problem recently through using stellar mass functions at different
environmental densities and redshifts up to $z \sim 1.4$ (Figure~3).  
By fitting the 
spectral energy distributions of galaxies in the BRIJK bands to different 
star formation histories, [5] calculated the most likely 
stellar mass to light (M/L) ratio for all galaxies within the DEEP2 
spectroscopic survey.  Through the use of the observed K-band
flux for these galaxies, and the derived stellar M/L ratio, a stellar
mass is calculated.  The mass function
for galaxies in high density environments compared to those in low density
environments, as measured through a
3rd nearest neighbor statistic in shown in Figure~3.  

Figure~3 shows that mass functions in low and high density 
environments are similar up to $z \sim 1.4$.  This implies that  
environment in a very broad sense is not a 
critical component for the formation of galaxies - that is, a galaxy or a 
potential proto-galaxy's formation by $z \sim 1.4$ does not strongly depend
upon whether it is in a dense region or a lower density region. There
are however some differences between mass functions in low and high
density environments, as galaxies at the extreme
ends of the environmental density distribution show
significant differences.  There are more
massive galaxies in higher density environments (cf. [36]), but the 
overall mass function shape, and the characteristic masses
of star formation galaxies (i.e., the downsizing) are similar.  
This furthermore implies that environmental processes that could trigger, or 
halt, star formation in groups are not major effects, at least since 
$z \sim 1.4$.  This is partially simply  another way of saying that most 
galaxy stellar mass is
formed by $z \sim 1.4$. However, there is some star formation evolution
in both field and group/cluster galaxies as observed through direct
measures [41], as well as increased
blue fractions (i.e., Butcher-Oemler).  However, this additional star
formation does not induce large amounts of new star formation in massive
galaxies, and most star formation is occurring in low-mass galaxies 
at $z < 1.4$ in both
low and high density environments.

\subsection{Young Galaxies at $z > 2$}
\label{sec:4}

Large samples of galaxy groups at redshifts larger than $z \sim
1.4$ do not yet exist.  What we do have is considerable evidence that 
$z > 2$ galaxies appear to be clustered, and are in environments that are 
either group-like, or forming into groups.  In fact, there is considerable
evidence that higher redshift galaxies are strongly clustered, with the
most massive galaxies (examples shown in Figure~4) the most clustered 
[28,20,39].

\begin{figure}
\centering
\includegraphics[height=5cm]{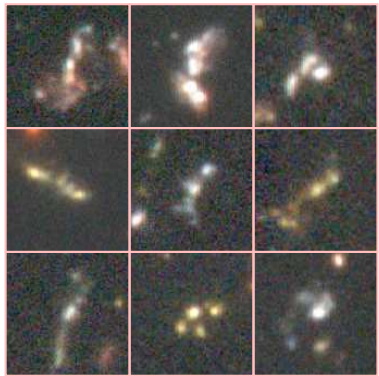}
%
%
\caption{Examples of high redshift bright galaxies as seen in the Hubble
Deep Field.  These types of galaxies are the most clustered and often
shown direct signs through color gradients, and structures, for a recent
merger origin.}
\label{fig:1}       
\end{figure}

Examples of this clustering can be seen all the way back to where
the earliest galaxies are seen. For example, deep Hubble
Space Telescope (HST) imaging of bright 
$z \sim 6$ QSOs show an excess of red $(i-z)$ galaxies surrounding these
systems [56].  If these QSOs are the sites of
massive galaxy progenitors, then it seems likely that they and their 
companion galaxies will evolve to become a virialized system - perhaps
a group or a cluster.  Making the connection between these
`groupings', and modern clusters/groups is not as difficult as it might
seem.  The large statistical excess of objects with red $(i-z)$ colors 
strongly suggests that these objects are spatially 
associated with each other.  They
are furthermore found within a projected radius of a Mpc or so - which is 
the typical size of a group or cluster core. 

Although we can detect galaxy clustering through various techniques out
to $z \sim 6$, often we know very little about the formation modes of
these galaxies, except that they are undergoing star formation.  Internal 
galaxy
properties as observed with HST, and integral field units, can only be studied 
in large numbers at $z < 3$.  There is now abundant information about these 
young galaxies which suggests that their
clustering, and perhaps grouping, is driving their formation and 
evolution.  

Galaxies at $z > 1.4$ are generally found to cluster
quite strongly  on small ($\sim 1$ Mpc)
scales.  A recent example of this 
comes from the Great Observatories Origins
Deep Survey (GOODS; 29).  Due to the high quality and
depth of the GOODS Hubble Space Telescope imaging it is 
possible to use the
Lyman drop-out technique to find galaxies undergoing unobscured star
formation at $z \sim 3$ and $z \sim 4$.  The correlation function  
for these systems has an excess at small scales over a power-law.  
This excess can be explained by a two
component model for galaxy clustering, such that a large scales galaxies
are in single halos, where as at small scales two galaxies per halo are
needed to
explain the excess [39].  There is also a luminosity dependence
to this clustering, such that the most luminous galaxies are the most
clustered [28,39].
We will address in the later part of this review how this
clustering might be driving the formation of galaxies found in 
early groups.

\section{Physical Processes in Groups}
\label{sec:3}

\subsection{Possible Formation Mechanisms}

Galaxy evolution occurs via both internal and external drivers.   The 
environments of galaxies should have some effect on how galaxies  
evolve.  For example, if galaxies are surrounded by other galaxies then 
gravitational 
interactions and mergers can induce the formation of new stars, remove
stars/gas from galaxies due to tides, increase the 
masses of galaxies through accretion of satellites, decrease the number of 
galaxies from mergers, and change the morphological types of galaxies.
Furthermore, if there exists an intragroup or intracluster medium then gas 
can be removed from member galaxies through processes such as 
ram-pressure stripping and strangulation [38].   All
of these processes are going on, but their strength and time-scales
are still very much debated.  We discuss each of these processes below, and 
why it appears that galaxy mergers are likely the dominate method by which 
galaxies in groups are evolving in the early universe ($z > 1.5$).

The efficiency of these various processes can be characterized by the 
velocity dispersion $\sigma$ of an environment.  Ram pressure stripping has an 
induced pressure force $\sim$ $\rho \sigma^{2}$, where $\rho$ is the 
density of the 
intergroup medium.  As the velocity dispersion of groups are typically 
$\sim 250-400$ \kms, the efficiency of this process in groups is likely not 
great, and 
its influence in clusters of galaxies where both $\rho$ and $\sigma$ are 
both high is also not clear.  Likewise high speed galaxy interactions,
such as galaxy harassment [46], 
are unlikely to produce significant effects in groups due to the 
low relative velocities of member
galaxies. This type of process is most efficient in rich clusters
of galaxies where the velocity dispersion is high, and galaxies are rapidly 
interacting with cluster members [45].  Interactions
between a group galaxy and the potential of its group [6] 
are also unlikely to be very effective given the low masses of groups.
  
Strangulation/starvation is the processes whereby hot gas is removed from a
galaxy's halo after it enters a hot medium.  This is proposed to 
halt the star formation in the accreted galaxy, as eventually no hot gas is 
left to cool and form new stars.  It is however not a fast process, and has a 
time-scale of
roughly a Gyr in clusters. It also requires a removal mechanism, which is 
usually an  intracluster medium, that interacts and removes hot gas from 
orbiting galaxies.   These mechanisms cannot be effective however
as we see star formation occurring in groups. Unless these groups are
young, the time-scale for truncation of star formation via ram-pressure
and strangulation must be longer than a Hubble time. This
leaves galaxy mergers and low velocity interactions, and non-gravitational 
processes, such as AGN feedback, as major effects that drive the 
evolution of group members.

\subsection{Galaxy Mergers}

Galaxy mergers should be common in groups of galaxies. This is due
to the low velocity dispersion of galaxies in groups, and the close 
proximity of members.  Various types of nearby groups, such as the 
compact groups and the fossil groups, are possibly the result of the
merger process.  A simple way to understand this is through dynamical 
friction effects which have a time-scale that varies roughly as 
as $\tau_{\rm merge} \sim \sigma^{3}$. Lower velocity dispersion groups 
therefore have a shorter time scale for mergers than galaxies in clusters.
This is one reason why galaxy-galaxy mergers in the centers of massive 
clusters are rare. Simple calculations show that groups 
with velocity dispersions $> 300$~\kms\, will not have a significant number of 
mergers over a Hubble time. This is one reason why galaxy mergers, 
almost by definition, must occur in groups of galaxies.
  
\begin{figure}
\centering
\hspace{-1cm}
\includegraphics[height=6.5cm]{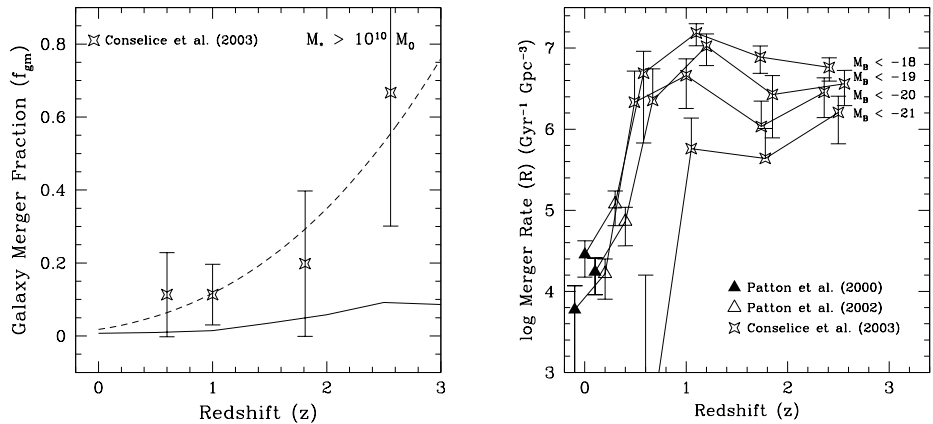}
%
%
\caption{Left panel: The galaxy merger fraction evolution for systems
with M$_{*} > 10^{10}$ \solm as a function of redshift [13].  Right panel: The galaxy merger rate in units of Gyr$^{-1}$ and Gpc$^{-3}$
as a function of redshift.  The merger rate at $z > 1.5$ is very high for 
galaxies at all luminosities.  The merger fraction for the most massive
galaxies is also high - around 50\% [19].}
\label{fig:1}       
\end{figure}

Galaxy mergers and interactions are sometimes observed in nearby groups, 
which may be a common way to induce star formation in these
systems.   However, galaxy mergers in nearby groups, and in groups
up to $z \sim 1$, are not expected to be common. Simulations
and analytical calculations of the dynamical friction process show that
galaxies in groups become more centrally
concentrated by about a factor of two since $z \sim 1$ [8], but do
not necessarily merge.  The effect is strongest for groups with lower 
velocity dispersions, $< 150$ \kms.  
[8] argues that merging in groups at $z \sim 0.4$
should be low, about 2\% per Gyr per group.
Observations tend to agree that the merger fraction for massive
galaxies is not high  (but does increase) at
redshifts $z \sim 0-1$  [13,42,43].

The merger rate for galaxies in groups should increase significantly 
at earlier times, that is at $z > 2$. The reason is simply because the 
universe was denser and galaxies were physically closer together, and should
thus merge more often. A simple argument shows this to be the case.  
In a ${\rm \Lambda}$-dominated universe, the 
mass density increases as the Hubble parameter squared, or 
H$^{2} \sim (1+z)^{3}$, with a merger rate 
$\sim \sqrt(H^{2}) \sim (1+z)^{1.5}$.  This merger rate explains why
we see a mix of galaxy types in groups, such as 
early types and spirals with bulges.  These spheroidal components were
formed when systems underwent mergers early in the universe, with 
disk/bulge systems the spheroids who were able to re-acquire a disk via
gas accretion from the intergalactic medium.

Finding mergers and calculating the merger rate at $z > 1$ has been carried
out using deep Hubble Space Telescope imaging [13].  The
merger fraction evolution out to $z \sim 3$ has been computed utilizing 
structural methods, such as the CAS system [11,15,16], and 
rest-frame optical observations of galaxies with NICMOS imaging of the 
Hubble Deep Field [22].  This imaging
revealed that galaxy morphology changes gradually from normal ellipticals
and spirals to peculiar galaxies at $z \sim 0$ to $z \sim 3$
[18].  In the nearby universe most of
the bright M$_{\rm B} < -20$ galaxies are normal Hubble types -- spirals
and ellipticals. However, when we view younger galaxies at higher redshifts
we find that the fraction, and number densities, of peculiars rises at the
expense of normal galaxies [47].  
This morphology-redshift relation has been
interpreted as an increase in the merger fraction with time [13].

The merger fraction varies with magnitude and stellar mass, such
that the brightest and most massive galaxies have the highest merger fractions
at $z \sim 2.5$.  This merger fraction declines steeply with redshift
approximately as a power-law, f$_{\rm m} \sim (1+z)^{3-5}$ (Figure~5). 
The merger rate can be calculated using 
N-body models of the merger process to obtain the time-scale
for mergers to occur as seen through the CAS system [19].   Using models with 
various orbital properties and
viewing angles, the time-scale for identifying mergers in the
CAS method is roughly 0.38$\pm$0.1 Gyr for galaxies
with stellar masses $>$~10$^{10}$ \solm.  Knowing this time-scale allows
us to calculate the merger rate evolution for galaxies as a function of
time and stellar mass. The result of this is shown in Figure~5.  

Integrating the
merger rate with time lets us determine the number of major mergers a galaxy
with a given initial stellar mass has undergone since $z \sim 3$. This simple 
calculation, explained in detail in Conselice (2006) [19], results 
in 4.4$^{+1.6}_{-0.9}$ 
major mergers since $z \sim 3$ for galaxies with initial masses 
$>$ 10$^{10}$~\solm.   This allows galaxies with stellar masses
of $\sim 10^{10}$ \solm, which tend to be among the most massive
galaxies at $z > 2$, to grow by a factor of 10-15 to contain as much stellar
mass as the most massive galaxies in the local universe.  Most of
this merger activity occurs at $z > 1.5$, with on average no mergers
occurring at $z < 1$ for the most massive systems.

What does an increase in the merger rate with redshift have to do
with galaxy evolution in groups?  We unfortunately do not yet know how
the merger rate varies with redshift {\em and} environment, but through
a chain of observations, we can argue that these mergers are occurring
in group-like environments.  The argument is simple - the mergers
we see occurring at $z > 2$ are in the most massive galaxies, which
previously must of been bound pairs or groups.
These massive galaxies are also clearly the most clustered,
as shown by [28] and [39].  The
scale of this excess clustering is small - about 1 Mpc - similar to the size
of a group. 

\section{Discussion}

\subsection{Environmental Correlations}

As we have already described in this review, there are several 
environmental correlations between the local density of a galaxy and that 
galaxy's properties.    Perhaps one of the most interesting 
characteristics  of these correlations is that there is little to no 
global environmental influence on the evolution of galaxies. This 
can be shown in a number of ways, including the fact that the star 
formation rate of galaxies, measured through H$\alpha$ equivalent widths, 
correlates with the local projected surface density of galaxies.  What is 
surprising is that this is independent of global environment, such that
the correlation is nearly the same  
in both clusters with $500 < \sigma < 1000$ \kms\, and within groups
with $\sigma < 500$ \kms\, [3]. 

What is the origin of local density sensitive properties of galaxies?  
This relates
to the classic `nature vs. nurture' debate about whether galaxy properties are 
imprinted early in their history, or whether they are transformed by
their environment.  We can argue now with some confidence that the 
observed correlation between galaxy properties and local density is likely
a `nature' process at low redshift, but a minor one that largely
involves halting star formation in accreted spirals. Except for
low mass systems [45], it is unlikely that environment 
induces significant morphological effects outside of mergers.  At higher
redshifts where most star formation and morphological evolution occurs,
the eventually properties of a galaxy are sealed by its local environment.

We can argue this through observational properties of galaxies and
basic theory. First, areas of higher density are predicted to
form earlier than those in lower density areas, as simple density
arguments and simulations show [51].  This implies
that the first galaxies should form in higher density areas.  This however
does not necessarily imply that galaxies in high density regions should be 
different (other than age) from galaxies in lower density areas.
Observationally, we know that the most massive galaxies at $z > 2$ tend
to have peculiar morphologies, and average stellar masses of $z \sim 10^{10}$
\solm\, [18,47].
Elliptical and spiral morphologies were not in place
until $z \sim 1-1.5$, and the morphology-density relation is already
in place by then [49].  The seeds of the morphology-density
relation therefore must have occurred even earlier when massive galaxies are
already highly clustered.

Galaxies in higher density regions in the early universe are rapidly 
merging with each other.  The progenitors of these galaxies
are not spirals, but simply post-mergers, with multiple mergers
occurring in a short time interval of a few Gyr [19],
providing no time for the establishment of a stable morphology.  Disk
galaxies are likewise able to form after the merger epoch has ended.
In this sense the morphology and stellar mass of a galaxy is set by the 
local density
in which it forms.  The global density is less important for
driving galaxy evolution, as merging activity is the dominant process.
By their nature, mergers are a local process driven by potentials dominated 
by a few massive
systems, or what we call in the local universe galaxy groups.  

The fact that the star forming properties of galaxies only depend upon
local environment, and not global environment, is a strong indicator that
whatever process is driving the decrease in star formation with local 
environment
is not related to the total velocity dispersion (or density) of the system 
where a galaxy is located [35, 3], nor to 
the density of the intracluster or intragroup medium.  The quenching of
star formation in massive galaxies is however perhaps not driven entirely
by environmental effects.   The star formation properties of a galaxy 
appear to correlate with galaxy mass more strongly than with environmental 
density. The internal properties of galaxies, either
through regulation with an active nucleus [33], or
from the time-scale for the exhaustion of gas must be responsible
for shutting down star formation.  What is not yet clear is why massive 
galaxies do not 
reestablish a cold gas supply, nor have their hot gas cool and form stars.  
This is perhaps related to the galaxy downsizing whereby star
formation is truncated in higher mass galaxies before lower mass
systems at $z < 1$ [5].

\subsection{Dwarfs and the Global Environment}

One galaxy property that does not simply correlate with local
environment is the faint end of the luminosity function, or the
ratio of dwarf to giant galaxies. This correlation is such that in higher 
density 
global environments, the number of low mass galaxies per giant galaxy is 
much higher than in lower density environments [25].  
This tells us first of all that galaxy clusters cannot form
through the mergers of lower mass galaxy groups.  This implies
that in {\em global} high density environments there are processes that
somehow produce low mass galaxies.  There is
no definitive agreement on how this occurs, with both a primordial origin
suggested [52], as well as scenarios in which dwarfs
are formed after clusters are in place [12,14]. For
a more detailed discussion of this issue see [17].

\section{Summary}
\label{sec:3}

Various observational techniques allow us to study the evolution of
galaxies in groups, and in group-like environments up to $z \sim 6$.  
By studying galaxies in groups at various 
redshifts we can determine the modes by which most galaxies evolved.
Four main features of galaxies in groups suggest how evolution has occurred
in this most common environment.
\vspace{0.2cm}

\noindent I. The morphological and star forming properties of galaxies in
low redshift galaxy groups 
reveal that multiple galaxy formation modes have occurred.    This is
due to the presence of star forming galaxies, such as spirals and irregulars, 
as well as
evolved galaxies, namely ellipticals and dwarf ellipticals.  There are also
several examples of galaxies evolving in nearby groups through
interactions/mergers.
\vspace{0.2cm}

\noindent II. Groups, in a traditional sense of having a measured velocity dispersion and
found within a small volume, can be identified out to $z \sim 1.4$.
Out to these redshifts we know that the group environment is very common
with as much as 50\% of all galaxies located in groups.
\vspace{0.2cm}

\noindent III. Galaxies in groups out to $z \sim 0.5$ evolve in a similar manner as the 
field, although groups tend to have a more evolved population at all      
redshifts thus far probed.  This implies that the galaxy formation process,
or at least residual star formation, is halted more quickly in groups
than in the field. This may however be an effect of groups
containing more massive galaxies, which end their star formation earlier
than low mass systems.
What triggers the star formation in galaxies in
groups is still not resolved, and the cause may not differ from
what is triggering star formation in field galaxies.
\vspace{0.2cm}

\noindent IV. At higher redshifts, there is evidence for groups in the form 
of  strong galaxy clustering, and merging, which produces larger
galaxies.  The universe was denser at high redshift by
a factor H$^{2}$, and thus  galaxies
were closer together.   The time-scale for these systems to merge
is fairly quick, and this is likely the method whereby most early
type galaxies and bulges were formed.  This is also a natural method for
putting the morphology-density relation into place.  What remains
a mystery is what causes the end to star formation in the most massive
galaxies, and why no further star formation occurs.
\vspace{0.2cm}

I thank the organizers of the ESO conference on galaxy groups for
their invitation to present this review, for their support, and 
their forbearance while it was written up.   I also thank Meghan
Gray for valuable comments which improved the presentation of this
review, and Brian Gerke for providing  Figure~2.
This work was supported by a National Science Foundation Astronomy
\& Astrophysics Fellowship and by PPARC.

%
%
%
%
%

%
%



\printindex
\end{document}